\documentclass[useAMS,usenatbib,usegraphicx]{mn2e}
\usepackage{amssymb}

\newcommand{\mnras}{MNRAS}
\newcommand{\aj}{AJ}
\newcommand{\aap}{A\&A}
\newcommand{\apj}{ApJ}
\newcommand{\apjl}{ApJ}
\newcommand{\apjs}{ApJS}
\newcommand{\prd}{Phys.\ Rev.\ D}
\newcommand{\physrep}{Physics Reports}

\title[Galaxy clustering in the gamma-ray background]{Imprint of
  galaxy clustering in the cosmic gamma-ray background}
\author[S.~Ando and V.~Pavlidou]{Shin'ichiro Ando\thanks{E-mail:
  ando@tapir.caltech.edu} and Vasiliki Pavlidou\thanks{E-mail:
  pavlidou@astro.caltech.edu}\thanks{Einstein (GLAST) Fellow}\\%
California Institute of Technology, Pasadena, CA 91125, USA}

\begin{document}

\date{Accepted 2009 August 24. Received 2009 August 24; in original
  form 2009 April 16}

\pagerange{\pageref{firstpage}--\pageref{lastpage}} \pubyear{2009}

\maketitle

\label{firstpage}

\begin{abstract}%
Star-forming galaxies are predicted to contribute considerably
 to the cosmic gamma-ray background (CGB) as they are confirmed
 $\gamma$-ray emitters and are the most numerous population of
 $\gamma$-ray sources, although individually faint.
Even though the {\it Fermi Gamma-ray Space Telescope} will be able to
resolve few star-forming galaxies individually, their fractional
 contribution to the CGB should become far more significant than it was
 for past measurements of the CGB as many of the brighter, formerly
 unresolved sources will be resolved out.
Thus, the clustering feature of galaxies imprinted on the CGB might be
 detectable by {\it Fermi}.
In anticipation of such measurements, we calculate the predicted angular
 auto-power and cross-power spectra of the CGB from normal galaxies.
We find that the amplitude of the auto-power spectrum is smaller than
 that for other sources such as blazars and dark-matter annihilation;
 the shape is also characteristic.
We also show that the cross-power spectrum with galaxy surveys features
 larger amplitude.
{\it Fermi} should be able to detect the correlation signature in both
 the auto-power and cross-power spectra at angular scales of
 $\sim$1--10$\degr$ after 5-yr of operation.
Such a detection would be valuable in confirming the level of the
 star-forming galaxy contribution to the CGB, and more importantly, in
 serving as a tool in the effort to discriminate between possible
 origins of the CGB.
\end{abstract}


\section{Introduction}

Star-forming galaxies are confirmed $\gamma$-ray sources.
They emit  $\gamma$-rays produced in hadronic interactions between
cosmic-ray nuclei and interstellar gas, and in leptonic interactions
between cosmic-ray electrons and secondaries with interstellar gas and
light \citep*[e.g.][]{Stecker1970, Stecker1973, FK1984, Dermer1986,
Strong2000}. 
Diffuse emission from the Milky Way is, in fact, the brightest
feature of the $\gamma$-ray sky, as demonstrated by SAS-2
\citep{SAS273}, COS-B \citep{COSB82}, the Energetic Gamma-Ray
Experimental Telescope (EGRET) onboard the {\it Compton Gamma-Ray
Observatory (CGRO)} \citep{Hunter1997}, and by the first-light results
of {\it Fermi Gamma-ray Space Telescope} \citep{Ritz2009}.
Other than the Milky Way, the only star-forming galaxy detected in
gamma-rays is the Large Magellanic Cloud \citep{Hartman1999}, because
normal star-forming galaxies are individually faint in $\gamma$-rays
\citep{Pavlidou2001}.
However star-forming galaxies are very numerous, and their collective
emission is likely to make a substantial contribution
(Pavlidou \& Fields 2002, hereafter PF02) 
to the cosmic gamma-ray background (CGB), measured
with EGRET \citep{Sreekumar1998}.

The Large Area Telescope (LAT) onboard {\it Fermi} will further refine
the CGB measurement with improved energy and angular resolutions,
whereas it will only detect no more than three additional galaxies as
individual sources \citep[Small Magellanic Cloud, M~31, and maybe
M~33;][]{Pavlidou2001}.
Therefore, normal galaxies would be a guaranteed source of the CGB for
{\it Fermi}-LAT, and their contribution would be at essentially the same
level as it was for EGRET.
Other contributors such as blazars, on the other hand, will be
substantially reduced with respect to their fractional contributions to
the EGRET CGB \citep{Stecker1999}, as the {\it Fermi}-LAT will resolve
many of them \citep[$\gtrsim$1000, depending on the luminosity
function; see, e.g.,][]{Narumoto2006, Dermer2007}.
Normal galaxies also have a characteristic spectral feature---a peak,
tracing the hadronic origin of their emission.
As a result, when the (more spectrally featureless) blazar
contribution is reduced, the contribution from normal galaxies to the
CGB could be dominant at energies around a few hundred MeV
(PF02),\footnote{Note that while PF02
found that the normal galaxies were likely to have a maximal
contribution to the total CGB energy flux at energies $\sim$1 GeV, this
result was a consequence of the ``GeV excess'' in the EGRET measurement
of the Milky-Way spectrum which recent {\it Fermi} observations in mid-Galactic-latitudes have not reproduced
(Johannesson \& Fermi LAT Collaboration 2009; \citealt{Abdo2009}) implying it was likely an instrumental effect. In this work,
we use a Milky-Way spectrum compatible with no GeV excess and find
the normal galaxy peak to reside at lower energies (see \S~\ref{intensity}).} which allows for an almost contamination-free set of
photons.

Galaxies are clustered following the large-scale matter distribution
in the Universe, and this clustering feature should be imprinted on
the CGB.
The anisotropy of the CGB has recently been studied theoretically by a
number of authors, in order to look for signatures of various
contributing sources such as blazars \citep{Ando2007a, Ando2007b},
galaxy clusters \citep{Ando2007a, Miniati2007}, type Ia supernovae
\citep{Zhang2004}, and dark-matter annihilation \citep*{Ando2006,
Ando2007b, Cuoco2007, Hooper2007, Cuoco2008, Siegal-Gaskins2008,
Lee2008, Taoso2008, Fornasa2009, Ando2009}.
The same approach should also be taken for the normal star-forming
galaxies.
Should this signature be detected in the {\it Fermi}-LAT data, it
would be extremely useful in a variety of ways:
\begin{enumerate}
\item As a consistency check.---%
If an energy range is identified spectrally where the normal-galaxy
contribution to the CGB is believed to be strongly dominant,  then the CGB photons in
this range {\em must} exhibit anisotropy properties consistent with our
understanding of normal-galaxy clustering.
\item As a powerful tool to disentangle multiple CGB components.---%
Instead of having the CGB strongly dominated by normal galaxies in some
energy range, an equally likely scenario is to have a balanced mixture
of normal galaxy photons and photons from different source classes.
In this case, as much information as possible is needed to disentangle
the different CGB contributions.
In this context, the angular power spectrum is as important a clue as
the shape of the energy spectrum of the contributions from different
populations (see, e.g., \citealp{JSGVP2009} on how the two can be combined
when information from the energy spectrum alone is insufficient to break
the degeneracy between different components).
The importance of the angular power spectrum in the case of normal
galaxies is further emphasized because their clustering properties are
very well constrained through galaxy surveys \citep[e.g.,][]{Cole2005,
Maller2005, Percival2007}.
\item As a complement to anisotropy studies of other populations.---%
As normal galaxies provide a guaranteed contribution to the CGB
for {\it Fermi}-LAT, the CGB anisotropy due to normal galaxies is
also a guaranteed background to any anisotropies studies using the
diffuse background.
For this reason, it is important to calculate the anisotropy properties
of the normal galaxy signal and understand its uncertainties and its
sensitivity to input parameters and assumptions.
\end{enumerate}

In this paper, we seek to calculate the expected angular correlation
of the CGB signature due to $\gamma$-ray emitting normal star-forming
galaxies.
We consider two quantities: the angular auto-power spectrum of the CGB
($C_\ell^{\gamma \gamma}$) and the angular cross-power spectrum between
the CGB map and some galaxy catalog ($C_\ell^{\gamma g}$).
An advantage of the auto-power spectrum analysis is that it can be
performed immediately after {\it Fermi}-LAT has obtained a sufficiently
deep all-sky $\gamma$-ray map, with $\gamma$-ray data alone.
As such, it does not suffer from uncertainties introduced through the
use of galaxy catalogs, such as issues of completeness and dust
corrections.
On the other hand, even given the additional uncertainties mentioned
above, taking the cross-correlation between the CGB map and a galaxy catalog
provides, as it turns out, a better way to detect the normal galaxy
angular signature, because of the large statistics of the large-scale
galaxy surveys.

This paper is structured as follows.
In \S~\ref{intensity}, we discuss the model we adopt to calculate the
contribution of normal star-forming galaxies to the $\gamma$-ray
background.
In \S~\ref{autospectrum}, we calculate the predicted angular
auto-power spectrum form star-forming galaxies, and in
\S~\ref{crossspectrum} we discuss the cross-correlation between the
normal-galaxy component of the CGB and galaxy catalogs.
We summarize and discuss our conclusions in \S~\ref{discussion}.

\section{Gamma-ray background from normal galaxies}\label{intensity}

We follow PF02 to derive a formulation for the mean CGB
intensity from normal galaxies.
We adopt their assumptions, and update our calculation with more recent
determinations of the cosmic star-formation history and of the
Milky-Way $\gamma$-ray spectrum.

The CGB intensity for photons with energy $E$ (in units of photon
number per unit area, time, solid angle, and energy range) is given by
\begin{equation}\label{iofe}
  I (E) = \frac{c}{4\pi} \int dz \frac{\dot n_{\gamma, \rm
    com} [(1+z)E, z]}{H(z)},
  \label{eq:I 2}
\end{equation}
where $\dot n_{\gamma, \rm com}$ is the comoving $\gamma$-ray emissivity
density, and $H(z)$ is the Hubble function.
We assume that the differential $\gamma$-ray luminosity (photons per
time per unit energy range) simply scales as star-formation rate
$\psi(z)$ and gas-mass fraction $\mu(z)$:
\begin{equation}
L_\gamma (E,z) = \frac{\psi(z)}{\psi_{\rm MW}} \frac{\mu(z)}{\mu_{\rm
 MW}}L_{\gamma, {\rm MW}} (E).
\end{equation}
where the quantities with the subscript ``MW'' represents those for
the Milky Way.
With the comoving number density of galaxies $n_{\rm gal}$, the
emissivity is then
\begin{equation}
\dot n_{\gamma, {\rm com}} (E,z) = L_\gamma n_{\rm gal}
= L_{\gamma, {\rm MW}}(E) \frac{\dot \rho_\ast (z)}{\psi_{\rm MW}}
\frac{\mu(z)}{\mu(0)},
\end{equation}
where $\dot \rho_\ast (z) \equiv \psi(z) n_{\rm gal}$ is the global
star-formation-rate density.\footnote{Here we have explicitly
assumed, unlike PF02, that most of the normal-galaxy
$\gamma$-ray emissivity at a certain redshift comes from galaxies with
similar $\gamma$-ray properties; it is the properties of that typical
galaxy that evolve with cosmic time according to the product of the
cosmic star-formation rate and the gas-mass fraction histories.}
Now, assuming that the sum of gas mass and star mass is constant in
the typical galaxy, the gas-mass fraction is simply given by
\begin{equation}
  \mu(z) = 1 - (1 - \mu_{\rm MW})
  \frac{\int_\infty^{z} dz (dt/dz) \dot \rho_\ast (z)}
  {\int_\infty^{0} dz (dt/dz) \dot \rho_\ast (z)}.
\end{equation}
Thus, given the cosmic history of the star-formation-rate density, one
could compute $\mu(z)$ by backwards de-evolving the present-day
Milky-Way gas mass fraction, $\mu_{\rm MW}$. The assumption of 
a total baryonic mass of galaxies staying constant in time is not 
necessarily realistic, as star formation is partly fueled by newly
accreted gas \citep{Prodanovic2008}. However, the overall effect of the details of the gas fraction evolution 
is relatively small ($\sim$factor of two, PF02). A more realistic modeling of the evolving gas fraction, including the effects of infall, will be addressed in an upcoming 
publication. 

For the present study, we adopt a model given by \citet{Hopkins2006}
for the global star-formation-rate density as a function of redshift,
$\dot \rho_\ast (z)$.
For the Milky-Way parameters, following PF02 and
references therein, we use $\psi_{\rm MW} = 3.2 M_{\sun}$ yr$^{-1}$ and
$\mu_{\rm MW} = 0.14$.
Lastly, we parametrize the Milky-Way $\gamma$-ray luminosity as
$L_{\gamma, {\rm MW}}(E) = 1.36 \times 10^{39}
(E/600~\mathrm{MeV})^{-\kappa}$ s$^{-1}$ MeV$^{-1}$, where $\kappa =
1.5$ for $E \le 600$ MeV and $\kappa = 2.7$ for $E > 600$ MeV.
This parametrization comes from a broken power-law fit to the
``GALPROP conventional'' \citep{Strong2004} model of the energy spectrum
of the diffuse Milky-Way $\gamma$-ray emission (which is compatible with
no GeV excess).\footnote{Note that this is not the latest version of
GALPROP that is used in the Fermi LAT data analysis and in the
comparison with Fermi LAT data on the diffuse emission from the
Galaxy. However, the anisotropy models we are discussing here are not
very sensitive to the details of the input single-galaxy intensity
spectrum. The aspects of the cumulative normal galaxy intensity spectrum
that affect our analysis the most are the energy of the peak, and the
energies above which we are in the power-law tail of the spectrum; both
these issues  can be adequately treated using  the simple models we
adopt here, and for this reason we do not engage in a more detailed
analysis of the cumulative intensity spectrum. We will return to the
latter in an upcoming publication.}  We set the normalization by
requiring that the energy integral of
$L_{\gamma,\rm MW}$ above 100~MeV is $2.85\times 10^{42} {\rm \,
photons \,\, s^{-1}}$ (see PF02 and references
therein).

\begin{figure}
\includegraphics[width=8.4cm]{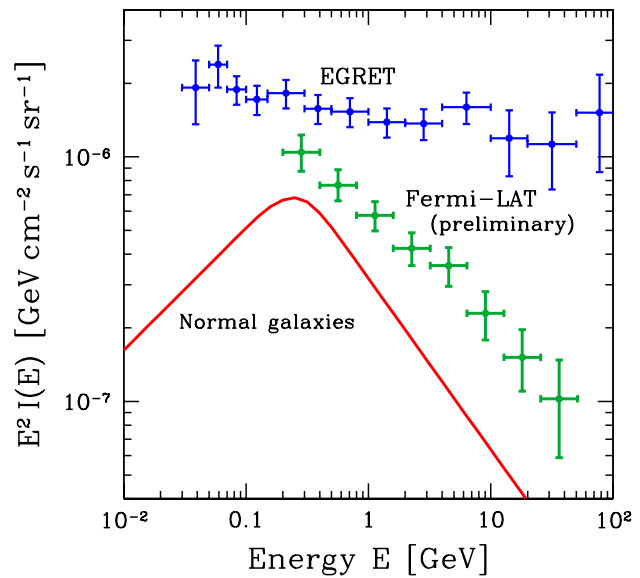}
\caption{The CGB intensity from normal galaxies, compared with the
 Sreekumar et al.\ (1998) determination of the CGB from  EGRET data and preliminary {\it Fermi} data.}
\label{fig:spectrum}
\end{figure}

In Fig.~\ref{fig:spectrum}, we plot, with the solid line, the $\gamma$-ray intensity $E^2
I(E)$ from normal galaxies, compared with the Sreekumar et al.\ (1998) determination 
of the CGB from EGRET data.
The galaxy contribution appears to be important in particular for
energies between 50 MeV and 1 GeV. We point out that due to the
shift of the spectral break in the Milky-Way diffuse emission spectrum
from 850 MeV (which was the location of the break in EGRET data which
suffered from the GeV excess) to 600 MeV (the location of the break in
GALPROP conventional),
the peak of the normal galaxy contribution correspondingly shifted from
$\sim$500 MeV in PF02 to $\sim$250 MeV in
Fig. \ref{fig:spectrum} in this work.
Additionally, the contribution of normal galaxies to the CGB declines
with energy above 1 GeV faster than it did in PF02, as
the high-energy slope of the Milky Way spectrum adopted here (2.7) is
steeper than the value implied by EGRET data (2.4) and adopted by
PF02. It is worth noting that a preliminary analysis of {\it Fermi} data
indicates that the slope of the CGB spectrum at high energies may be
substantially steeper (consistent with $\sim E^{-2.45}$, see
M. Ackermann for the LAT
Colaboration\footnote{http://www-conf.slac.stanford.edu/tevpa09/ \\
Ackermann090714v2.ppt}) than the EGRET measurement. To illustrate this
point, in Fig. \ref{fig:spectrum}, we also plot the preliminary {\it
Fermi} CGB results.

\begin{figure}
\includegraphics[width=8.4cm]{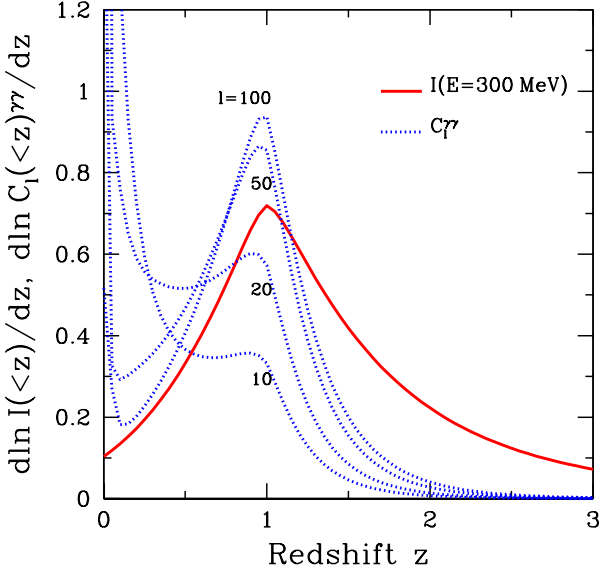}
\caption{Contribution from unit redshift range to the mean CGB
  intensity at 300 MeV (solid) and angular auto-power spectrum at $\ell
  = 10$, 20, 50 and 100 (dotted).}
\label{fig:z_dist}
\end{figure}

In Fig.~\ref{fig:z_dist}, we plot, with the solid line, the integrand of Eq. (\ref{iofe}) in
units of the integral as a function of redshift at $E = 300$ MeV; this quantity represents the contribution to the mean CGB
intensity at a given energy from galaxies in a specific redshift range.
Following the evolution of the cosmic star-formation rate, it peaks at
$z \simeq 1$ and declines for higher redshifts.

\section{Angular auto-power spectrum for the CGB from normal galaxies}\label{autospectrum}

The angular auto-power spectrum of the CGB map due to normal galaxies
is given by
\begin{equation}\label{Cgammagamma}
C_\ell^{\gamma\gamma} 
= \frac{c}{16\pi^2 I^2(E)} \int dz \frac{\dot n_{\gamma, {\rm
 com}}^2 ([1+z]E, z)}{H(z) r^2} P_{\rm gal}\left(\frac{\ell}{r},
 z\right),
\label{eq:angular auto-power spectrum}
\end{equation}
where $r$ is the comoving distance and $P_{\rm gal} (k, z)$ is the
galaxy power spectrum at comoving wave number $k$ and redshift $z$
\citep[e.g.,][]{Ando2007b}.
The multipole $\ell$ corresponds roughly to the angular scale of
$\theta = 180\degr / \ell$.
Note that we defined $C_\ell^{\gamma\gamma}$ as the variance of
the fluctuation from the mean intensity in units of steradian.

The galaxy power spectrum is a well measured quantity according to the
modern galaxy surveys \citep[e.g.,][]{Cole2005, Maller2005,
Percival2007}.
It traces the underlying matter power spectrum.
To compute the latter, we adopt the halo-model approach
\citep{Seljak2000, Cooray2002} with the linear transfer function given
by \citet{Eisenstein1999}, which gives a reasonable fit to the galaxy
power spectrum with a moderate correction for the bias, e.g., $b_{\rm
gal} = 1.11$ \citep*{Afshordi2004}.

\begin{figure}
\includegraphics[width=8.4cm]{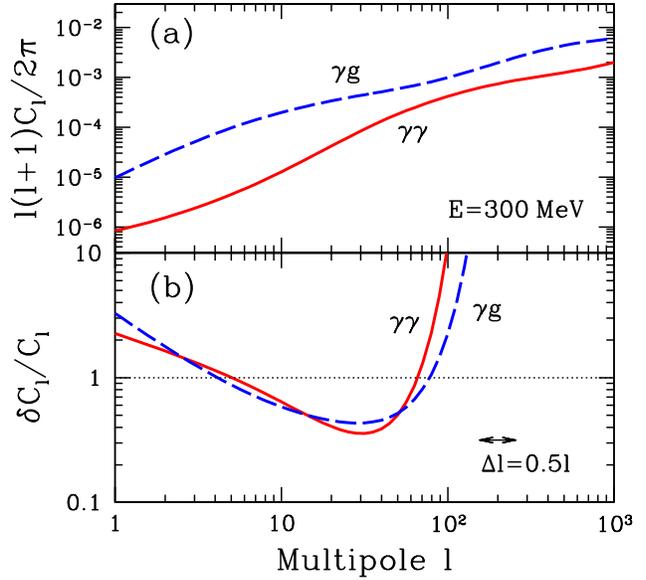}
\caption{(a) The angular auto-power ($\gamma\gamma$; solid) and
  cross-power ($\gamma g$; dashed) spectra for $E = 300$ MeV.  The
  cross-correlation is taken with a 2MASS-like galaxy catalog.  (b)
  The relative errors of the power spectra after 5-yr all-sky
  measurement with {\it Fermi}-LAT.  The bin width, $\Delta \ell = 0.5
  \ell$, is shown as the arrowed line.}
\label{fig:C_l}
\end{figure}

In Fig.~\ref{fig:C_l}(a), we show the angular auto-power spectrum
$\ell (\ell + 1) C_\ell^{\gamma \gamma} / 2\pi$ for $E = 300$ MeV, as a
function of multipole $\ell$.
In the multipole range between 1 and 10$^3$, $\ell(\ell+1)
C_\ell^{\gamma \gamma} / 2 \pi$ ranges from $10^{-6}$ to $10^{-3}$,
which is much smaller than the case of other sources.
For instance, in the case of blazars, $\ell (\ell + 1)
C_\ell^{\gamma\gamma} / 2\pi$ would be no smaller than $\sim$10$^{-4}$
even at large angular scales.
In the case of dark-matter annihilation in the extragalactic halos, it
could be as large as 0.1 at $\ell = 10^3$ \citep{Ando2006, Ando2007b},
or even larger in the case of annihilation in the Milky-Way subhalos
\citep{Siegal-Gaskins2008, Ando2009}.
In addition to the amplitude, the shape of the power spectrum might also
serve as a diagnostic as it is also different for different source
populations.

\begin{figure}
\includegraphics[width=8.4cm]{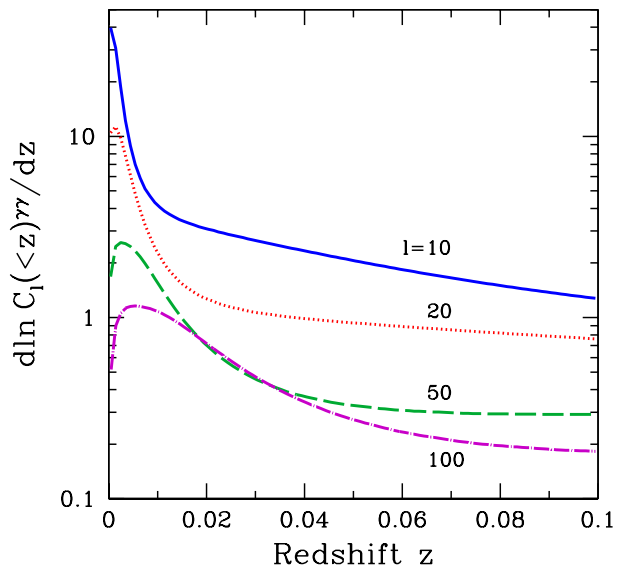}
\caption{The same as Fig.~\ref{fig:z_dist}, focused on low-redshift
 range.}
\label{fig:z_dist_lowz}
\end{figure}

In Fig.~\ref{fig:z_dist}, we also show the contribution to the angular
auto-power spectrum from a given redshift range, i.e., $d\ln
C_\ell^{\gamma \gamma} / dz$ (the integrand of Eq. (\ref{Cgammagamma})
as a function of redshift in units of the integral) for multipoles $\ell
= 10$, 20, 50, and 100, which, as we show below, are the observationally
relevant scales.
For large angular scales, e.g., $\ell = 10$ and 20, the dominant
contribution comes from low redshifts mainly because of the $r^{-2}$
dependence of the integrand in Eq.~(\ref{eq:angular auto-power
spectrum}).
For smaller angular scales, on the other hand, the distribution develops
a second peak at $z = 1$, reflecting the dependence on the cosmic
star-formation rate.
Therefore, in principle, we can probe different redshift ranges by
observing $C_\ell^{\gamma \gamma}$ for various $\ell$, even though
these quantities are obtained after the redshift information is
integrated out.
Note that the relative contributions of different redshift ranges to
$C_\ell^{\gamma\gamma}$ are also different from those to the mean
intensity $I(E)$.
In Fig.~\ref{fig:z_dist_lowz}, we show the same redshift distribution
for $C_\ell^{\gamma\gamma}$, but focusing on the lower-redshift range
$z<0.1$.
Towards larger distances, correlations of galaxies are averaged out
quickly, and this effect is more prominent for large angular-scale
modes as expected.

We now examine the auto-correlation detectability with the {\it
Fermi}-LAT.
The $1\sigma$ errors for $C_\ell^{\gamma\gamma}$ measurements are
given by
\begin{equation}
\delta C_\ell^{\gamma \gamma} = \sqrt{\frac{2}{(2\ell+1) \Delta \ell
f_{\rm sky}}}
\left(C_l^{\gamma \gamma} + \frac{C_P + C_N}{W_\ell^2}\right),
\label{eq:error for auto-power}
\end{equation}
where $f_{\rm sky} = \Omega_{\rm sky} / 4 \pi$ is the fraction of the
sky measured, $\Delta \ell$ is the bin width for which we use $0.5
\ell$, and $W_\ell = \exp (-\ell^2 \sigma_b^2 / 2)$ is the window
function with the angular resolution $\sigma_b \approx 1\fdg 2$ for 300
MeV photon.
The first term represents the cosmic variance, and the second  the
shot noise due to finite statistics of galaxy ($C_P$) and photon
($C_N$) counts.
The Poisson noise due to galaxies $C_P$ is obtained by
Eq.~(\ref{eq:angular auto-power spectrum}) with replacement $P_{\rm
gal} (k,z) \to n_{\rm gal}^{-1}$; for the comoving number density of
galaxies $n_{\rm gal}$, we use 10$^{-2}$ Mpc$^{-3}$ and assume it is
independent of redshifts.
We thus obtain $C_P = 3.4 \times 10^{-8}$ sr.
The Poisson noise due to finite photon count is given by $C_N =
\Omega_{\rm sky} / N_\gamma$, where $N_\gamma$ is the number of
photons received from $\Omega_{\rm sky}$.
We estimate $N_\gamma = E I(E) A_{\rm eff} T_{\rm eff} \Omega_{\rm
sky} \approx 5.0 \times 10^6 f_{\rm sky}$, where we used $E = 300$ MeV,
$A_{\rm eff} = 6000$ cm$^2$, $T_{\rm eff} = T \Omega_{\rm fov} / 4
\pi$, $\Omega_{\rm fov} = 2.4$ sr (LAT field of view), and assumed
a 5-yr all sky survey ($T = 5$ yr).
Thus, we obtain $C_N = 2.5 \times 10^{-6}$ sr, which dominates the noise
term due to finite galaxy counts $C_P$. Note that the uncertainties calculated 
here are conservative, as we have not included the effects of smearing 
with energy within the assumed energy bin ($\Delta E \sim E$, with the 
bin extending from $E$ to $2E$). Although the $C_\ell$ are independent of energy as long as we
are in the power-law tail of the intensity spectrum, the angular resolution  
$\sigma_b$, and thus the uncertainties $\delta C_\ell$, do depend on energy. However, 
since in the spectrum high-energy tail the intensity is decreasing with energy as $\sim E^{-2.7}$, 
the photons in the energy bin will be dominated by the low-energy photons, and the effect of 
energy smearing will be small. In addition, as $\sigma_b$ {\em decreases} with increasing energy, 
the inclusion of higher-energy photons will result in a decrease of the overall uncertainty. 

In Fig.~\ref{fig:C_l}(b) we plot, with the solid line, $\delta C_\ell^{\gamma\gamma} /
C_\ell^{\gamma\gamma}$ assuming all-sky coverage ($f_{\rm sky} = 1$).
There appears to be a sweet spot between $\ell \approx 5$ and 70,
where one can claim positive detection of galaxy clustering in the CGB
with 5-yr {\it Fermi} data.
Below this region, as we have only $2\ell + 1$ modes for fixed $\ell$,
 $C_\ell^{\gamma\gamma}$ cannot be constrained very well (cosmic
variance).
For $\ell$'s larger than 100, corresponding to $\theta \lesssim
1\degr$, the errors become exponentially large because of the limited angular
resolution of {\it Fermi}-LAT.

\section{Cross-correlation with galaxy catalog}\label{crossspectrum}

We now discuss the cross-correlation between the CGB map and the existing
galaxy catalogs.
The angular cross-power spectrum is given by
\begin{eqnarray}
C_\ell^{\gamma g}
&=& \frac{1}{4\pi I(E) N_g}\int dz \frac{\dot n_{\gamma, {\rm com}}
([1+z]E, z)}{r^2} \frac{dN_g}{dz}
\nonumber\\&&{}\times
P_{\rm gal}\left(\frac{l}{r},z\right),
\label{eq:angular cross-power spectrum}
\end{eqnarray}
where we define $dN_g/dz$ as the redshift distribution of galaxies and
$N_g$ is the total number of galaxies of the catalog:
\begin{equation}
N_g = \int dz \frac{dN_g}{dz}.
\end{equation}
The galaxy auto-power spectrum can also be computed with these
quantities as
\begin{equation}
C_\ell^{gg}
= \frac{1}{c N_g^2} \int dz \frac{H(z)}{r^2}
\left(\frac{dN_g}{dz}\right)^2
P_{\rm gal}\left(\frac{l}{r},z\right).
\label{eq:galaxy auto-power}
\end{equation}

\begin{figure}
\includegraphics[width=8.4cm]{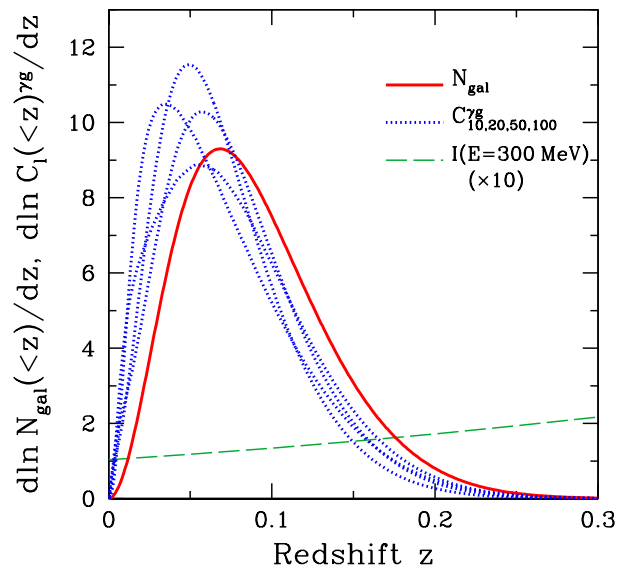}
\caption{Redshift distribution of a 2MASS-like galaxy catalog (solid)
  and angular cross-power spectrum at $\ell = 10$, 20, 50, and 100
  (dotted).  The redshift dependence of $10\times$ the mean CGB intensity 
is shown for  comparison (dashed).}
\label{fig:z_dist_cross}
\end{figure}

For the present study, we consider a galaxy catalog similar to the Two
Micron All Sky Survey (2MASS) Extended Source Catalog
\citep{Jarrett2000}.
This is a full sky ($f_{\rm sky} \sim 1$), near infrared survey of
galaxies whose median redshift is around $z \sim 0.1$ and total number
is $N_g \sim 10^6$.
The redshift distribution $d\ln N_g / dz$ is shown as a solid curve in
Fig.~\ref{fig:z_dist_cross}, for which we used fitting formula given
in \citet{Afshordi2004}.

Using this galaxy catalog and the CGB emissivity at $E = 300$ MeV, we
compute the angular cross-power spectrum $\ell (\ell + 1) C_\ell^{\gamma
g} / 2\pi$, showing it as a dashed curve in Fig.~\ref{fig:C_l}(a).
The amplitude of the cross-power is larger than that of auto-power by
about an order of magnitude, which would make the former easier to be
detected.
In Fig.~\ref{fig:z_dist_cross}, we show contributions from unit
redshift ranges to $C_\ell^{\gamma g}$ for $\ell = 10$, 20, 50, and
100 as dotted curves.
Unlike the case of auto-power spectrum,  the redshift
distribution is fairly similar for different angular scales, because
the galaxy distribution $dN_g / dz$ has a much sharper peak than
$dI/dz$.

The $1\sigma$ errors of the cross-power spectrum is estimated by
\citep[e.g.,][]{Zhang2004, Cuoco2007}
\begin{eqnarray}
\delta C_\ell^{\gamma g}
&=& 
\sqrt{\frac{1}{(2\ell+1)\Delta \ell f_{\rm sky}}}
\biggl[(C_\ell^{\gamma g})^2
\nonumber\\&&{}
+\left(C_\ell^{\gamma\gamma} +
\frac{C_P+C_N}{W_\ell^2}\right)
\left(C_\ell^{gg} + C_{N,g} \right)\biggr]^{1/2},
\end{eqnarray}
where we use Eq.~(\ref{eq:galaxy auto-power}) for $C_\ell^{gg}$ in
this expression; $C_{N,g} = \Omega_{\rm sky} / N_g = 1.5 \times
10^{-5} f_{\rm sky}$ sr is the galaxy shot noise.
The errors for $C_\ell^{\gamma g}$ are plotted as a dashed curve in
Fig.~\ref{fig:C_l}(b), which shows similar prospects to the case of
auto-power spectrum, for detecting the galaxy clustering in the CGB
anisotropy.
The sweet spot is slightly wider than that for the auto-power spectrum.

\section{Discussion and conclusions}\label{discussion}

In the calculations for both the mean intensity and anisotropy, we
assumed that the $\gamma$-ray emissivity of all galaxies at the same
redshift is the same, rescaled from the emissivity of the Milky Way
using the  cosmic star-formation rate as well as the gas-mass fraction.
This implicitly assumes that all the galaxies of interest are Milky-Way-like in
their $\gamma$-ray properties.
Although this is clearly not true for all galaxies,  what it really
amounts to is assuming that {\em most $\gamma$-ray photons} emitted by star-forming galaxies come from Milky-Way--like, properly de-evolved sources.
This in turn is not an unreasonable assumption.
Milky-Way-like objects
are rich in both star formation and gas, so they are expected to be the
most $\gamma$-ray bright among normal star-forming galaxies of the same
epoch \citep[see, e.g.,][]{Pavlidou2001}.
This is the reasoning behind taking, as a first approximation, all normal galaxies contributing to the CGB to have a single luminosity at a given redshift,  instead of
using a luminosity function.

A notable exception to this general rule is that of starburst galaxies, which, 
depending on the details of cosmic ray confinement, could individually be one to two 
orders of magnitude brighter in $\gamma$-rays than the typical Milky-Way--like galaxy of the 
same cosmic epoch, as well as have harder energy spectra at high energies \citep*[see, e.g.,][]{Thompson2007}. However starburst galaxies would be best treated as a distinct source 
class as far as their anisotropy properties are concerned---compared to normal star-forming 
galaxies, the population of starburst galaxies consists of few and bright sources, and the 
anisotropy at small angular scales could be considerably stronger, even if the 
overall contribution of starbursts to the CGB is lower. We will return to this issue in a future publication. 

\begin{figure}
\includegraphics[width=8.4cm]{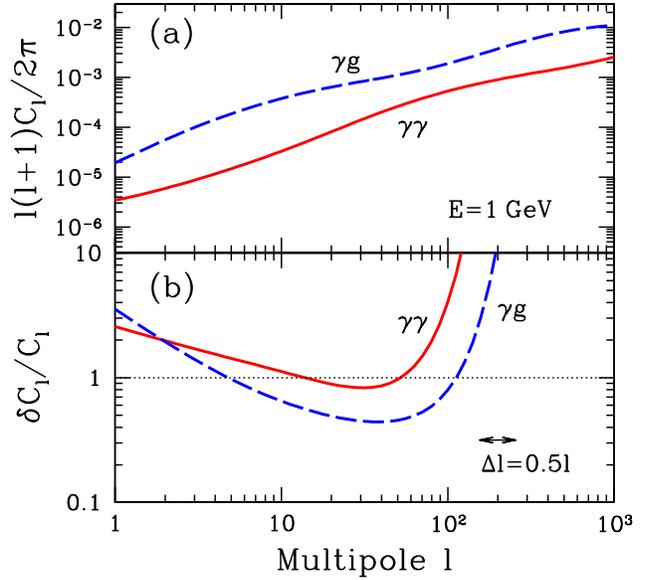}
\caption{Same as Fig.~\ref{fig:C_l}, for photons of $E = 1$ GeV.}
\label{fig:C_l_1GeV}
\end{figure}

Until this point, we have concentrated on photons of $E = 300$ MeV, as the $\gamma$-ray energy flux
spectrum due to normal galaxies peaks at this energy (see
Fig.~\ref{fig:spectrum}).
We now also discuss the results for higher energy photons $E = 1$ GeV.
At this energy, although the energy flux from star-forming galaxies is lower, the LAT effective area increases to 8000 cm$^2$, and also the
angular resolution improves to $\sim$0$\fdg$8 \citep{lat09}.
In Fig.~\ref{fig:C_l_1GeV}, we show the same plots as Fig.~\ref{fig:C_l}
but for $E = 1$ GeV photons.
We find that especially for the cross-correlation, the detection prospects
are still pretty good even though the number of photons received would
decrease.
In addition, the multipole range for the detection becomes larger, above
$\ell = 100$ for $C_\ell^{\gamma g}$.

Although we fixed the galaxy bias to be $b_{\rm gal} = 1.11$
throughout \citep{Afshordi2004}, the relevant value of the bias might
be different.
This is because while $b_{\rm gal}$ refers to all galaxies, we are not interested in elliptical galaxies, faint dwarfs, or gas-poor dwarfs,  which do not emit significant amount of $\gamma$-rays.
Thus, the bias of star-forming $\gamma$-ray-bright galaxies might
differ from that of all the galaxies as inferred
from galaxy catalogs.
However, our results are not very sensitive to the value of $b_{\rm gal}$ as long as the 
true value is not significantly different from 1.

The emission from individual galaxies we have considered is entirely due
to the interaction between cosmic rays and interstellar gas and light;
any contribution from point sources within galaxies has not been
accounted for. However, using $\gamma$-ray observations of the Milky Way
for guidance, we expect that the contribution of point sources to the total
gamma-ray emission of a star-forming galaxy is small. First, the
relative intensity of the diffuse flux is much higher than the total
emission due to resolved point sources in the Milky Way \citep[see,
e.g.,][]{Hartman1999}. Second, the good agreement between the Milky Way
diffuse emission as measured by {\it Fermi} LAT and GALPROP, indicates
that in the Milky Way point sources are not a dominant component in the
diffuse emission,  at least at mid-Galactic latitudes.
The situation can be very different for early-type galaxies with little
star formation, in which most $\gamma$-ray emission would arise from
nonthermal processes in older populations of stellar remnants, such as
millisecond pulsars. However the contribution of early-type galaxies to
the CGB is expected to be very small.

We also comment on the nonlinear part of the galaxy power spectrum.
The angular scales where such nonlinearity becomes important are $\ell
\approx 10^3$ for $C_\ell^{\gamma\gamma}$ and $\ell \approx 100$ for
$C_\ell^{\gamma g}$.
Remembering that the angular resolution of {\it Fermi}-LAT for 1 GeV
photons corresponds roughly to $\ell \approx 100$, the nonlinear part
of the power spectrum does not affect the relevant result much.
The reason why nonlinearity becomes important at lower $\ell$ for the
cross-power is that with the 2MASS-like galaxy catalog the
contribution is biased to lower redshifts (compare redshift
distribution in Figs.~\ref{fig:z_dist} and \ref{fig:z_dist_cross})
that correspond to larger spatial scales for fixed angular scales
(note $k = \ell / r$ in the argument of galaxy power spectrum $P_{\rm
gal}$).

In conclusion, motivated by the fact that normal galaxies provide a
guaranteed contribution to the CGB and that {\it Fermi}-LAT has a good
sensitivity to measure it, we have theoretically computed the CGB angular
power spectrum due to normal galaxies.
We have calculated both the auto-power ($C_\ell^{\gamma\gamma}$) and
the cross-power ($C_\ell^{\gamma g}$) spectra, using the well measured
galaxy power spectrum; for the cross-power, we correlated the CGB with
a 2MASS-like galaxy catalog.
We found that the amplitude of $C_\ell^{\gamma\gamma}$ is smaller than
that for other sources such as blazars and dark-matter annihilation.
Still, {\it Fermi}-LAT can measure the significant feature of the
galaxy clustering for the multipole range $10 \lesssim \ell \lesssim
100$ in about 5 years.
The amplitude of the cross-power spectrum $C_\ell^{\gamma g}$ is
larger, and the detection prospects are better for higher-energy
photons.
We also found that the redshift ranges that contribute to the power
spectrum the most are different from the case of mean intensity.
This feature  might be helpful in probing
the $\gamma$-ray luminosity density from normal galaxies at various
redshift ranges.

\section*{Acknowledgments}

We thank John Beacom and Troy Porter for insightful comments that helped improve this paper. 
This work was supported by Sherman Fairchild Foundation (SA) and by NASA
through the GLAST Fellowship Program, NASA Cooperative Agreement:
NNG06DO90A (VP).

\label{lastpage}

\end{document}